\documentclass[fleqn,usenatbib]{mnras}

\usepackage{newtxtext,newtxmath}

\usepackage[T1]{fontenc}

\DeclareRobustCommand{\VAN}[3]{#2}
\let\VANthebibliography\thebibliography
\def\thebibliography{\DeclareRobustCommand{\VAN}[3]{##3}\VANthebibliography}

\usepackage{graphicx}	
\usepackage{amsmath}	
\usepackage{orcidlink}

\title[ML Cluster Classification with PHANGS-HST]{Star Cluster Classification using Deep Transfer Learning with PHANGS-HST}

\author[Hannon et al.]{Stephen Hannon \orcidlink{0000-0001-9628-8958},$^{1,18}$\thanks{E-mail: hannon@mpia.de}
Bradley C. Whitmore \orcidlink{0000-0002-3784-7032},$^{2}$
Janice C. Lee \orcidlink{0000-0002-2278-9407},$^{3,4,2}$
David A. Thilker \orcidlink{0000-0002-8528-7340},$^{5}$
\newauthor
Sinan Deger \orcidlink{0000-0003-1943-723X},$^{6,23}$
E.~A. Huerta \orcidlink{0000-0002-9682-3604},$^{7,8,9}$
Wei Wei,$^{9}$
Bahram Mobasher \orcidlink{0000-0001-5846-4404},$^{1}$
Ralf Klessen \orcidlink{0000-0002-0560-3172},$^{10,11}$
\newauthor
M\'ed\'eric Boquien \orcidlink{0000-0003-0946-6176},$^{12}$
Daniel~A.~Dale \orcidlink{0000-0002-5782-9093},$^{13}$
M\'elanie Chevance \orcidlink{0000-0002-5635-5180},$^{10,14}$
Kathryn Grasha \orcidlink{0000-0002-3247-5321},$^{15,16}$
\newauthor
Patricia Sanchez-Blazquez \orcidlink{0000-0003-0651-0098},$^{17}$
Thomas Williams \orcidlink{0000-0002-0012-2142},$^{18}$
Fabian Scheuermann \orcidlink{0000-0003-2707-4678},$^{19}$
Brent Groves \orcidlink{0000-0002-9768-0246},$^{20,21}$
\newauthor
Hwihyun Kim \orcidlink{0000-0003-4770-688X}, $^{3}$
J.~M.~Diederik Kruijssen \orcidlink{0000-0002-8804-0212},$^{14,22}$
and the PHANGS-HST Team
\\
$^{1}$Department of Physics $\And$ Astronomy, University of California, Riverside, CA, 92507, USA\\
$^{2}$Space Telescope Science Institute, Baltimore, MD, 21218, USA\\
$^{3}$Gemini Observatory/NSF’s NOIRLab, 950 N. Cherry Avenue, Tucson, AZ, 85719, USA\\
$^{4}$Steward Observatory, University of Arizona, Tucson, AZ 85721, USA\\
$^{5}$Department of Physics and Astronomy, The Johns Hopkins University, Baltimore, MD 21218, USA\\
$^{6}$TAPIR, California Institute of Technology, Pasadena, CA 91125, USA\\
$^{7}$Data Science and Learning Division, Argonne National Laboratory, Lemont, Illinois 60439, USA\\
$^{8}$Department of Computer Science, University of Chicago, Chicago, Illinois 60637, USA\\
$^{9}$Department of Physics, University of Illinois at Urbana-Champaign, Urbana, Illinois 61801, USA\\
$^{10}$Zentrum f{\"u}r Astronomie der Universit{\"a}t Heidelberg, Institut f{\"u}r Theoretische Astrophysik, D-69120 Heidelberg, Germany\\ 
$^{11}$Universit{\"a}t Heidelberg, Interdiszipli{\"a}res Zentrum f{\"u}r Wissenschaftliches Rechnen, D-69120 Heidelberg, Germany\\ 
$^{12}$Instituto de Alta Investigaci\'on, Universidad de Tarapac\'a, Casilla 7D, Arica, Chile\\
$^{13}$Department of Physics \& Astronomy, University of Wyoming, Laramie, WY, 82071, USA\\
$^{14}$Cosmic Origins Of Life (COOL) Research DAO, coolresearch.io\\
$^{15}$Research School of Astronomy and Astrophysics, Australian National University, Canberra, ACT 2611, Australia; kathryn.grasha@anu.edu.au\\
$^{16}$ARC Centre of Excellence for All Sky Astrophysics in 3 Dimensions (ASTRO 3D), Australia\\
$^{17}$Departamento de Física de la Tierra y Astrof\'isica, UCM, 28040 Madrid, Spain\\
$^{18}$Max Planck Institut f{\"u}r Astronomie, K{\"o}nigstuhl17, 69117 Heidelberg, Germany\\
$^{19}$Astronomisches Rechen-Institut, Zentrum f{\"u}r Astronomie der Universit{\"a}t Heidelberg, M{\"o}nchofstra{\ss}e 12-14, D-69120 Heidelberg, Germany\\
$^{20}$International Centre for Radio Astronomy Research, The University of Western Australia, 7 Fairway, Crawley, WA 6009, Australia\\
$^{21}$Research School of Astronomy and Astrophysics, Australian National University, Mt. Stromlo Observatory, Weston Creek, ACT 2611, Australia\\
$^{22}$Technical University of Munich, School of Engineering and Design, Department of Aerospace and Geodesy, Chair of Remote Sensing Technology, \\\hspace{2.2mm}Arcisstr. 21, 80333 Munich, Germany\\
$^{23}$The Oskar Klein Centre for Cosmoparticle Physics, Department of Physics, Stockholm University, AlbaNova, Stockholm, SE-106 91, Sweden
}

\date{Accepted XXX. Received YYY; in original form ZZZ}

\pubyear{2021}

\begin{document}
\label{firstpage}
\pagerange{\pageref{firstpage}--\pageref{lastpage}}
\maketitle

\begin{abstract}
Currently available star cluster catalogues from \textit{HST} imaging of nearby galaxies heavily rely on visual inspection and classification of candidate clusters. The time-consuming nature of this process has limited the production of reliable catalogues and thus also post-observation analysis. To address this problem, deep transfer learning has recently been used to create neural network models which accurately classify star cluster morphologies at production scale for nearby spiral galaxies ($D \lesssim 20$~Mpc). Here, we use \textit{HST} UV-optical imaging of over 20,000 sources in 23 galaxies from the Physics at High Angular Resolution in Nearby GalaxieS (PHANGS) survey to train and evaluate two new sets of models: i) distance-dependent models, based on cluster candidates binned by galaxy distance (9--12~Mpc, 14--18~Mpc, 18--24~Mpc), and ii) distance-independent models, based on the combined sample of candidates from all galaxies. We find that the overall accuracy of both sets of models is comparable to previous automated star cluster classification studies ($\sim$60--80 per cent) and show improvement by a factor of two in classifying asymmetric and multi-peaked clusters from PHANGS-HST. Somewhat surprisingly, while we observe a weak negative correlation between model accuracy and galactic distance, we find that training separate models for the three distance bins does not significantly improve classification accuracy. We also evaluate model accuracy as a function of cluster properties such as brightness, colour, and SED-fit age. Based on the success of these experiments, our models will provide classifications for the full set of PHANGS-HST candidate clusters (N$\sim$200,000) for public release.



\end{abstract}

\begin{keywords}
galaxies : star clusters : general
\end{keywords}



\section{Introduction}

The evolution of star clusters is inherently linked to the evolution of their host galaxies. Most star formation occurs in clustered regions within giant molecular clouds \citep{LADA03}, so star clusters and associations represent a fundamental unit in tracing the overall star formation cycle, which in turn informs us of the larger-scale dynamical evolution of galaxies \citep{ADAMO20}. 

Samples of star clusters and their ensemble properties have thus served as the basis for many studies that seek to better understand these processes (see \citealt{ADAMO20} for a review). For example, such studies inform us about the formation of globular cluster systems \citep{WHITMORE93,WHITMORE95,LARSEN01,KRUIJSSEN14}, the characterisation of the star cluster luminosity function \citep{WHITMORE99,LARSEN02} and the initial cluster mass function \citep{LARSEN09,CHANDAR10,CHANDAR16,MESSA18A,MESSA18B}, the spatial distribution of clusters and their hierarchical formation \citep{BASTIAN05B,GRASHA17,ELMEGREEN20}, correlations with various galactic parameters such as surface brightness, morphological type \citep{LARSEN99,LARSEN00}, star formation history \citep{BASTIAN05A}, and the timescales for the clearing of the natal gas of stars \citep{WHITMORE11,HOLLYHEAD15,GRASHA18,GRASHA19,HANNON19,MESSA21,HANNON22,CALZETTI23}, among many others. Physics at High Angular resolution in Nearby GalaxieS (PHANGS\footnote{\url{https://www.phangs.org}}; see PHANGS-HST, \citealt{LEE21}; PHANGS-ALMA, \citealt{LEROY21}; PHANGS-MUSE, \citealt{EMSELLEM21}; PHANGS-JWST, \citealt{LEE23}) represents one of the newest and largest extragalactic surveys to systematically study these topics addressing the complete star formation cycle on the cluster scale across a broad range of galactic environments.

As these star cluster studies have evolved toward survey scales, the size of their cluster samples has grown dramatically, with PHANGS-HST and the Legacy ExtraGalactic UV Survey (LEGUS; \citealt{CALZETTI15}), each containing tens of thousands of cluster candidates. In these surveys, clusters are categorised according to a four-class system based on morphology (\citealt{ADAMO17}; see Section~\ref{sec:Data} for class definitions), which not only crucially differentiates clusters from artefacts, but has also shown correlations with the physical properties of star clusters, including age and mass (e.g. \citealt{GRASHA15,GRASHA17,WHITMORE21,DEGER21}). The classification of these objects has historically been performed by one or more humans, which is time-consuming and thus effectively limits the sample size and depth. 

Recently, however, there has been exploration in the use of machine learning techniques for the rapid, automated, production-scale classification of star cluster candidates. \citet{GRASHA19} created a generally successful classification model ($\sim$70\% agreement with human classifications) using a bagged decision tree algorithm with star clusters from LEGUS \citep{CALZETTI15}, however it did not perform as well for more distant objects or for compact associations of stars. \citet{WEI20} and \citet{PEREZ21} then improved on these models by utilising deep learning with even larger samples of LEGUS galaxies, resulting in $\sim$10x greater recovery rates for compact associations. The agreement between deep learning models and human classifiers in particular ($\sim$70\% overall) rivals the consistency found between human classifiers, and thus highlights the viability of machine learning in producing cluster catalogues much more efficiently. 

While these models perform well for samples of objects from LEGUS, on which they were trained, they do not perform as well for the more recent cluster sample from PHANGS-HST. \citet{WEI20} and \citet{WHITMORE21} use LEGUS-trained models to classify PHANGS-HST objects from NGC 1559 and NGC 1566, respectively, and find a 10--20\% decrease in recovery rate relative to LEGUS samples for asymmetric clusters (Class 2) and compact associations (Class 3). One explanation for this is that in PHANGS-HST, the definition of Class 3 objects is more explicitly specified in order to avoid stellar pairs or triplets, which are sometimes categorised as Class 2 or 3 objects by LEGUS \citep{WHITMORE21}. Another possible explanation is that the PHANGS-HST cluster sample occupies a different distance range (4--24~Mpc, median = 15.5~Mpc; \citealt{ANAND20}) than the sample used for the LEGUS-based models (3--10~Mpc, median = 6.7~Mpc; \citealt{CALZETTI15}), and more distant objects have been shown to be associated with lower model accuracy (e.g., \citealt{PEREZ21}).

The present study aims to investigate these issues with two primary experiments. First, we will train a new set of deep learning models based on the PHANGS-HST sample of star clusters, available for 23/38 galaxies at the time of the writing of this paper. This will allow us not only to compare performance with previous models and various properties of the host galaxies, but also potentially provide a more accurate machine learning-based classification for the PHANGS-HST cluster catalogues. Secondly, we will experiment with 
distance-dependent models, in which separate models are trained on cluster candidates binned by galaxy distance (9--12~Mpc, 14--18~Mpc, 18--24~Mpc).

The organisation of this paper is as follows. Section~\ref{sec:Data} introduces the data used in this study, including the cluster sample, the classification system, and the inputs of the model. Section~\ref{sec:PriorModels} examines the accuracy of the current ML-based classifications used by PHANGS-HST. Section~\ref{sec:Experiments} details the experiments and procedure of the current study, while Section~\ref{sec:Results} presents the primary results of our model performance. Section~\ref{sec:Trends} examines additional correlations with model performance, and Section~\ref{sec:Summary} provides a summary of this study, as well as our general conclusions.

\section{PHANGS-HST Clusters}
\label{sec:Data}

The PHANGS-HST survey \citep{LEE21} consists of 38 spiral galaxies at distances of 4--24 Mpc \citep{ANAND20}, observed with the Hubble Space Telescope in five broad bands ($NUV-U-B-V-I$), using either WFC3/UVIS or archival ACS/WFC images. Critically, we are able to effectively resolve star clusters at these distances. The $\sim$0.04 $\arcsec$/pixel scale of \textit{HST} imaging can resolve objects with sizes between 1.7--9.0 pc across the distance range of the PHANGS-HST galaxies\footnote{The point source function of WFC3/UVIS at 5000\AA  is 0.067$\arcsec$. The size of a source as measured by the broadening of its image can be measured down to $\sim$0.2 pixels \citep{CHANDAR16,RYON17,BROWN21}, which corresponds to size limits of 0.5-1.3pc for the distance range in our sample.}, which is consistent with the effective radii of compact star clusters (typically 0.5--10 pc; \citealt{PORTEGIES10,RYON17,BROWN21}).

These sources are initially detected using \texttt{DOLPHOT} \citep{dolphin2000} and are then photometrically selected as cluster candidates based on the measurement of multiple concentration indices and the use of model star clusters \citep{THILKER22}. Co-author Brad Whitmore (BCW) then finalises our sample by visually inspecting the brighter candidates ($m_V$ $\lesssim$ 24~mag) to categorise each according to their morphology. Fainter candidates are visually inspected and categorised on an ad hoc basis to evaluate the performance of the models presented in this work (see Section~\ref{sec:clusterbrightness}). Following previous studies such as \citet{GRASHA15}, \citet{ADAMO17}, and \citet{COOK19}, there are four primary morphological classes, which are displayed in Figure~\ref{fig:ClusterClasses} and defined as follows: 

\begin{figure*}
\includegraphics[width =1.0\textwidth]{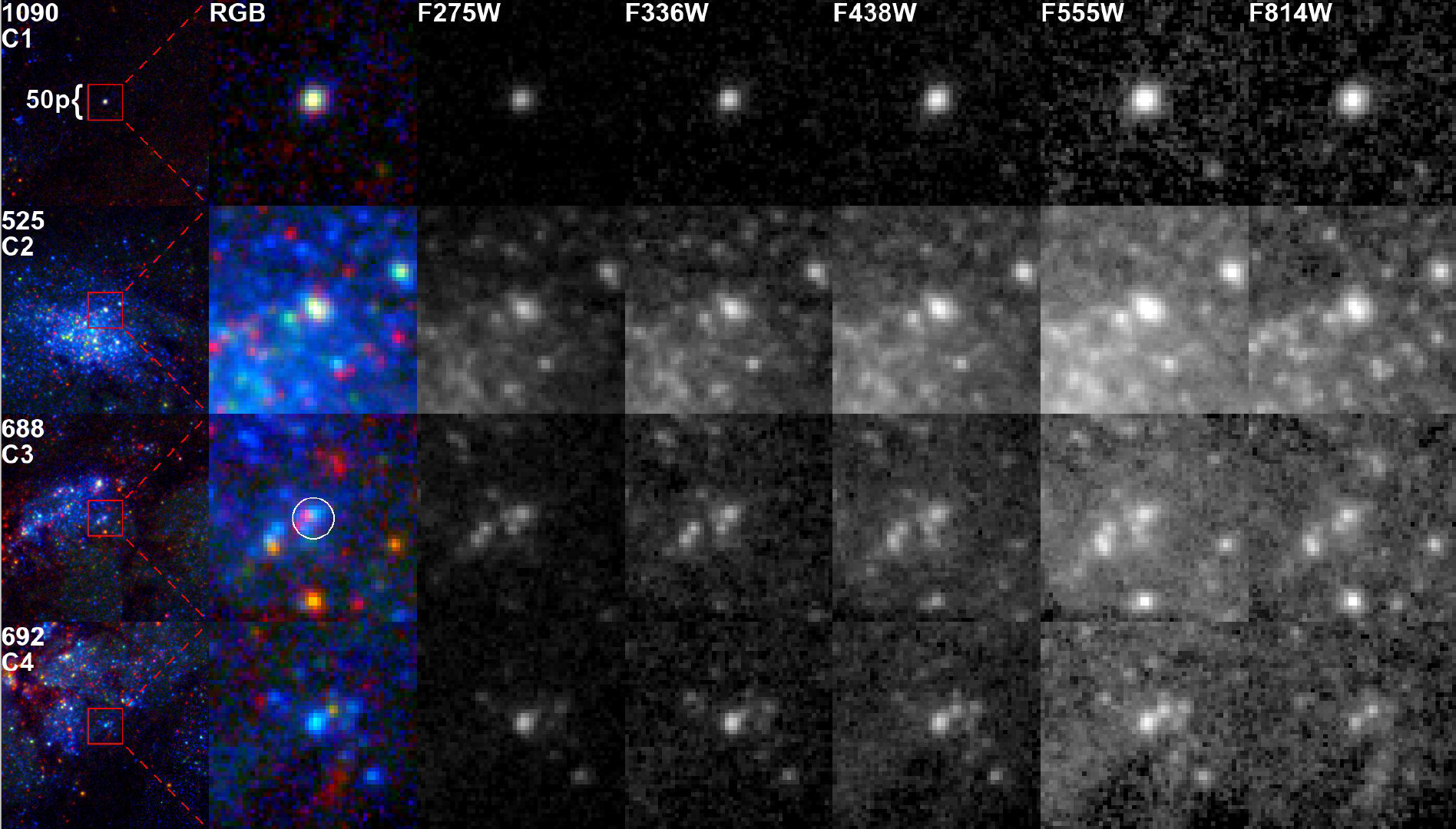}
\caption[RGB images of each of the four classes of star cluster morphology]{Examples of each of the four classes of training objects (Section~\ref{sec:Data}) from a medium-distance galaxy within our sample (NGC~4321; $D \sim 15$ Mpc). A 50 x 50 pixel region, centred on each target object (Section~\ref{sec:Data}), is extracted from each of the five photometric bands and stored in a single file to be used for training. RGB images (R = F814W, G = F438W + F555W, B = F275W + F336W) are provided in the left two columns along with each Object ID (top left) and morphological class (C1 = Class 1, e.g.) for reference.}
\label{fig:ClusterClasses}
\end{figure*}


\begin{enumerate}
    \item Class 1: compact, symmetric, single central peak, radial profile more extended relative to point source
    
    \item Class 2: compact, asymmetric, or non-circular (e.g., elongated), single central peak
    
    \item Class 3: asymmetric, multiple peaks, sometimes superimposed on diffuse extended source
    
    \item Class 4: not a star cluster (image artefacts, background galaxies, single stars, pairs and multiple stars in crowded regions)
\end{enumerate}

While these general definitions are used for both LEGUS and PHANGS-HST cluster samples, the Class~3 definition is further specified in PHANGS-HST to require at least four stars within a 5-pixel radius. This can be seen for Object~688 in Figure~\ref{fig:ClusterClasses}, where at least two bluer and two redder stars are visible within the five-pixel radius (denoted by the white circle) in the RGB image. This is a key change which eliminates pairs (e.g., Object~692 in Figure~\ref{fig:ClusterClasses}) and triplets, which have a higher probability of being chance superpositions of individual stars within crowded regions. As discussed later, the effect of this change is reflected in the accuracy of the LEGUS-based models (Section~\ref{sec:PriorModels}) and the PHANGS-based models (Section~\ref{sec:Results}) in the classification of PHANGS-HST cluster candidates. Additionally, ambiguities among these Class 3 objects, although they represent a significant fraction of young clusters (Section~\ref{sec:IndividualTrends}), often leads studies to limit their star cluster samples to Class 1 and 2 objects. To this end, PHANGS-HST employs a different strategy to identify Class 3-like objects as multi-scale stellar associations \citep{LEE21, LARSON22}.

At the time of the analysis for this paper, visual classifications had been completed for cluster candidates in 23 galaxies (24 fields), providing a sample of over 20,000 objects. 
Table~\ref{tab:sample} lists all of the galaxies used in this study along with the number of clusters in each morphological class, sorted by galaxy distance. Further information on the PHANGS-HST survey, the data processing pipeline, and the production of the star cluster catalogues can be found in \citet{LEE21}. 

\begin{table*}
\begin{tabular}{llrrrr}
\multicolumn{6}{c}{\textbf{PHANGS Clusters}} \\
\hline
\textbf{Field}                    & \textbf{Distance (Mpc)} & \textbf{Class 1} & \textbf{Class 2} & \textbf{Class 3} & \textbf{Class 4} \\
\hline
IC 5332*                           & 9.01                    & 78               & 154              & 148              & 233              \\
NGC 628E                          & 9.84                    & 51               & 41               & 22               & 98               \\
NGC 628C                          & 9.84                    & 264              & 225              & 189              & 582              \\
NGC 3351                           & 9.96                    & 140              & 177              & 173              & 1033             \\
NGC 3627                           & 11.32                   & 462              & 312              & 184              & 546              \\
\hline
{\textbf{Bin 1 Totals}}       & \textbf{}               & \textbf{995}     & \textbf{909}     & \textbf{716}     & \textbf{2492}    \\
\hline
\hline
NGC 5248                           & 14.87                   & 211              & 324              & 195              & 420              \\
NGC 4571                           & 14.9                    & 61               & 101              & 100              & 200              \\
NGC 4298*                          & 14.92                   & 173              & 103              & 79               & 188              \\
NGC 4689*                          & 15                      & 130              & 214              & 166              & 265              \\
NGC 1433                           & 15.17                   & 90               & 104              & 99               & 363              \\
NGC 4321                           & 15.21                   & 436              & 279              & 235              & 421              \\
NGC 4569                           & 15.76                   & 214              & 213              & 100              & 197              \\
NGC 4535                           & 15.77                   & 202              & 203              & 127              & 435              \\
NGC 1792                           & 16.2                    & 265              & 302              & 108              & 550              \\
NGC 4548                           & 16.22                   & 96               & 99               & 76               & 142              \\
NGC 4303                           & 16.99                   & 264              & 293              & 140              & 484              \\
NGC 1566                           & 17.69                   & 394              & 291              & 166              & 911              \\
\hline
{\textbf{Bin 2 Totals}}       & \textbf{}               & \textbf{2536}    & \textbf{2526}    & \textbf{1591}    & \textbf{4576}    \\
\hline
\hline
NGC 7496*                          & 18.72                   & 105              & 158              & 110              & 245              \\
NGC 1672                           & 19.4                    & 238              & 134              & 121              & 891              \\
NGC 1559                           & 19.44                   & 420              & 303              & 218              & 641              \\
NGC 1365                           & 19.57                   & 366              & 269              & 154              & 714              \\
NGC 685*                          & 19.94                   & 111              & 194              & 173              & 222              \\
NGC 4654                           & 21.98                   & 256              & 360              & 243              & 409              \\
NGC 2775                           & 23.15                   & 136              & 160              & 110              & 222              \\
\hline 
{\textbf{Bin 3 Totals}}       & \textbf{}               & \textbf{1632}    & \textbf{1578}    & \textbf{1129}    & \textbf{3344}    \\
\hline
{\textbf{Full Sample Totals}} & \textbf{}               & \textbf{4566}    & \textbf{4190}    & \textbf{2760}    & \textbf{9259}   \\
\hline
\end{tabular}
\caption[Number of objects in each of the four morphological classes for the 23 PHANGS-HST galaxies (in 24 fields), as classified by BCW]{Number of objects in each of the four morphological classes for the 24 PHANGS-HST fields, as classified by BCW. The total number of clusters within each of the three distance bins is also given, as well as the totals for the ``full" sample. Asterisks indicate galaxies which are not included in the training of the initial, distance-independent models. Distances from \citet{ANAND20} are listed.}
\label{tab:sample}
\end{table*}

\section{Accuracy of Prior Models}
\label{sec:PriorModels}

The PHANGS-HST project has produced cluster catalogues complete with cluster classifications determined by a star cluster expert (BCW) for 23 galaxies (24 fields). Along with these human-determined cluster classifications, each of the cluster candidates has morphological classifications predicted by 20 independently trained models presented in \citet{WEI20}, 10 of which were created using the \texttt{VGG19-BN} neural network architecture \citep{SIMONYAN14}, while the other 10 were created with \texttt{ResNet18} (\citealt{HE16}; Section~\ref{sec:Procedure} further details these architectures). As these models were trained on clusters from the LEGUS sample (5,147 total objects; see Table 1 of \citealt{WEI20}), we will henceforth refer to them as the ``LEGUS-based" models.

To produce these machine learning classifications, postage stamps of each of the objects in the PHANGS-HST cluster catalogues are first created as described in Section~\ref{sec:TrainingImages}. The full sample of objects is then fed through a single model for evaluation, the product of which is a list of predicted classes for all of the objects. The evaluation is repeated for each of the 20 LEGUS-based models, resulting in 10 \texttt{VGG19-BN} classifications and 10 \texttt{ResNet18} classifications. The mode class from each of the two neural network architectures is then chosen to represent the final classification for each object. These models, along with a Jupyter Notebook tutorial on how to use them to predict classifications for new catalogues of objects, have been made publicly available by the PHANGS-HST team via MAST (\url{https://archive.stsci.edu/hlsp/phangs-hst}).

Notably, the LEGUS-based models were trained on slightly different objects than those which make up the PHANGS-HST sample. First, as noted in Section~\ref{sec:Data}, the definition of a Class~3 object in PHANGS-HST has been further specified to eliminate pairs and triplets, which are sometimes included as Class~3 objects in LEGUS. Secondly, the LEGUS-based models were trained on a sample of objects that span a nearer galactic distance range ($\sim 3-10$~Mpc) than the PHANGS-HST sample examined in this study ($\sim 9-24$~Mpc). Finally, we differ in our strategy with non-detections (background-subtracted flux < 1 $\sigma$ for PHANGS-HST; \citealt{THILKER22}). The LEGUS-based models only used sources that were detected in at least four filters, and also set all pixel values to zero in any filter where the source was not detected. Our models, on the other hand, require detection in only three filters, and also retain the data in filters where non-detections are recorded. It is also important to note that the LEGUS-based models discussed in this work are the BCW models presented in \citet{WEI20}, which, like our models, are based on objects classified by the same, individual expert, BCW. By comparing the human-verified classes of PHANGS-HST objects with the classes predicted by the LEGUS-based models, we can examine the robustness of the models to these important distinctions. 


\begin{figure*}
\includegraphics[width =0.8\textwidth]{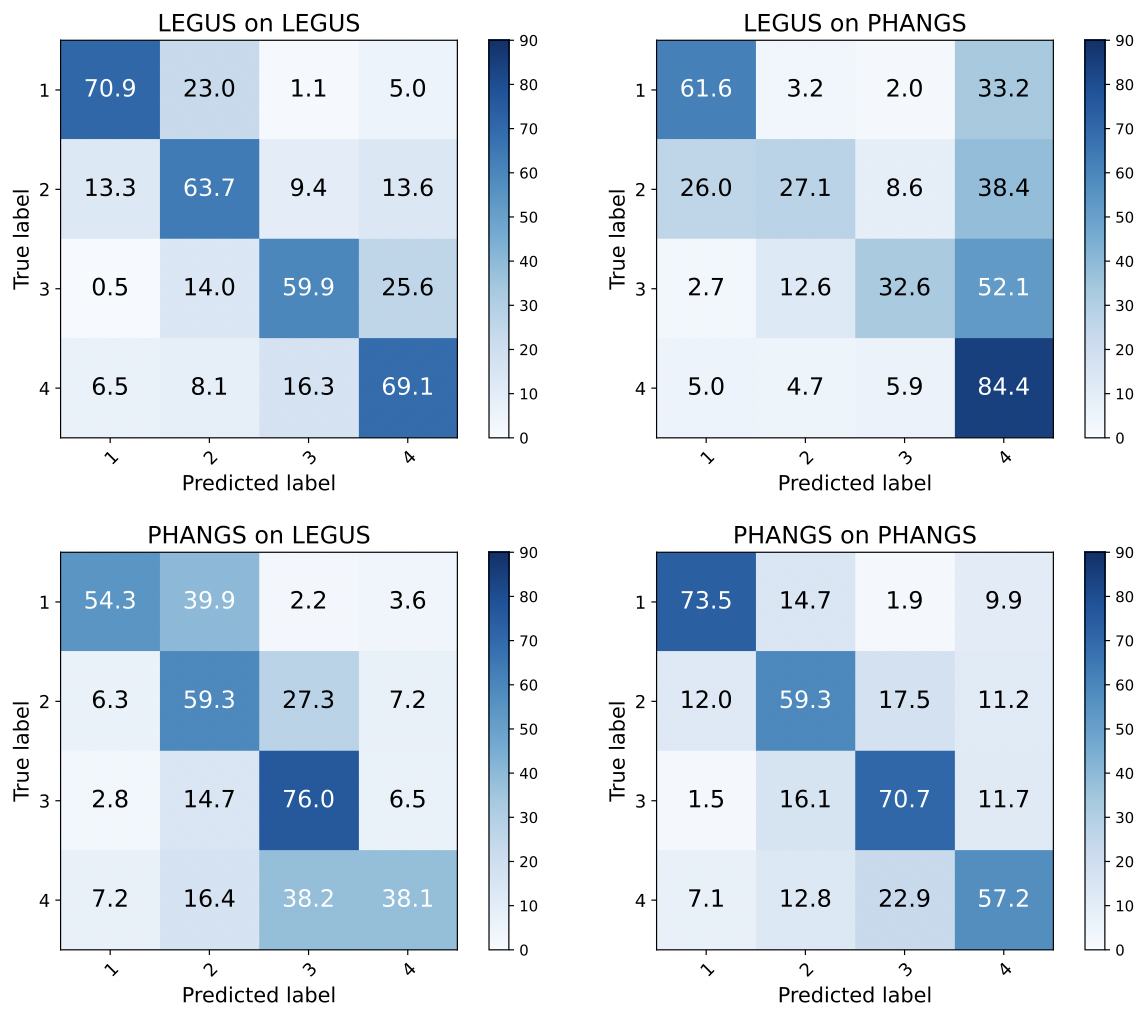}
\caption[Comparison of classification accuracies for the LEGUS-based and PHANGS-based models]{Comparison of classification accuracies for the LEGUS-based \citep{WEI20} and PHANGS-based models, with the model-determined labels and human-determined labels on the x-axes and y-axes, respectively. The top-left and top-right confusion matrices display the accuracy of the LEGUS-based models in classifying the LEGUS and PHANGS-HST candidate clusters, respectively. The bottom-left and bottom-right matrices display the accuracy of the PHANGS-based models in classifying LEGUS and PHANGS-HST candidates, respectively. While the LEGUS-based and PHANGS-based models classify objects from their respective samples with similar accuracy ($\sim$60-70\%), the LEGUS-based models are notably poor at correctly classifying Class~2 and 3 PHANGS-HST objects. Note that the results presented here are specific to the \texttt{VGG19-BN} models from each study, however, similar results are found for the \texttt{ResNet18} models which are discussed in the text.}
\label{fig:Wei_results}
\end{figure*}

Over the sample of $\sim$20,000 objects used in this study, we find that the LEGUS-based models predict Class 1 and 4 objects from the PHANGS-HST sample with reasonable accuracy; however, they do not accurately classify Class 2 and 3 objects. The agreement between BCW's classes ('True Label') and the machine learning classes ('Predicted Label') are displayed in the confusion matrices of Figure~\ref{fig:Wei_results}. The diagonal of each matrix represents the 1:1 agreement between the class determined by BCW and the class determined by the particular model. The top-left matrix displays the accuracy of the LEGUS-based models tested on LEGUS objects \citep{WEI20}. The top-right matrix displays the accuracy of those same LEGUS-based models when classifying objects in the PHANGS-HST sample. As shown in the upper-left cell of this confusion matrix, $\sim$60\% of the PHANGS-HST objects identified by BCW as a Class 1 cluster have also been given a Class 1 label by the LEGUS-based models. Following the diagonals downward, we find that Class 2, 3, and 4 objects show classification agreement for $\sim$30\%, $\sim$30\%, and $\sim$80\% of PHANGS-HST objects, respectively. 

For Class 2 and 3 objects, these percentages indicate poor agreement compared to the results of \citet{WEI20}. The recovery rate of Class 2 and 3 objects is reduced by half when the LEGUS-based models are applied to the PHANGS-HST objects, while we observe a more modest decrease in the recovery rate of the Class 1 PHANGS-HST objects (10\%). If we take a simple average of the accuracies for cluster objects (Class 1, 2, and 3), the LEGUS-based models drop from 65\% accuracy when classifying LEGUS clusters to 40\% for PHANGS-HST clusters.

If these models are to be used for production-scale cluster classifications, it is imperative that they can reliably delineate between cluster (Class 1, 2, 3) and non-cluster (Class 4), however we do not find this to be the case with the LEGUS-based models. We find that a significant fraction of Class 1, 2, and 3 PHANGS-HST objects are misclassified by the LEGUS-based models as non-clusters (Class 4). As shown in the rightmost column of the upper-right matrix, ~30--50\% of BCW Class 1, 2, and 3 objects have been misclassified by the models as Class 4. This is likely related to the observed good agreement for Class 4 objects (84\%), which may be explained by the fact that the more-distant PHANGS-HST objects and nearby background sources appear less-resolved than those found in the LEGUS sample.

Thus, we find that the current LEGUS-based models used to produce machine learning-based classifications for the PHANGS-HST sample of clusters do not achieve a sufficient level of accuracy, which motivates this study in training new machine learning models. Our results, shown in the bottom-right matrix of Figure~\ref{fig:Wei_results}, will be discussed in Section~\ref{sec:Results}.

\section{Training Experiments}
\label{sec:Experiments}


Considering the relatively poor performance of the LEGUS-based models in classifying PHANGS-HST clusters, we perform two experiments seeking to improve the reliability of such machine learning-based classification models for the PHANGS-HST sample of objects and potentially future star cluster samples.

For a direct comparison with the LEGUS-based models, we first train the neural networks using the full available sample of PHANGS-HST objects, which we refer to as our ``distance-independent" models. At the time of this particular training, cluster catalogues with human-verified classifications were available for 18 galaxies (19 fields). These galaxies span a distance range of $\sim 9-24$ Mpc and consist of 20,775 objects. 

To help illuminate whether the poor performance of the LEGUS-based models is due to the difference in galaxy distances in the samples, we divide our PHANGS-HST sample into three separate galaxy distance bins (9--12~Mpc, 14--18~Mpc, and 18--24~Mpc) and train the neural networks individually on each of the three samples. These models will be referred to as our ``distance-dependent" models. The bins for these were determined based on the most natural breaks found in the galaxy distances in the sample: $\Delta D$ = 3.55~Mpc between NGC 3627 and NGC 5248, and $\Delta D$ = 1.03~Mpc between NGC 1566 and NGC 7496. Because this training was performed subsequent to the distance-independent training, BCW classifications of cluster candidates were made available for five additional galaxies in this distance-dependent experiment, for a total of 23 galaxies (24 fields).  The 9--12~Mpc, 14--18~Mpc, and 18--24~Mpc distance bins consist of 5,112, 11,229, and 7,683 objects, respectively. Therefore, our smallest sample is comparable in size to the sample used for the BCW models from \citet{WEI20}, which consisted of 5,147 objects. 

Eighty percent of the objects in each sample are randomly chosen for their respective training set to be used for the machine learning process (Section~\ref{sec:Procedure}), while the remaining 20\% of objects are reserved as a validation set to evaluate the accuracy of the resultant models. 

The complete list of galaxies along with each of their human-verified cluster populations for these two experiments is displayed in Table~\ref{tab:sample}. 

\subsection{Training Set}
\label{sec:TrainingImages}

In order to perform deep transfer learning, we need to supply the neural networks with a set of images on which to train. For our dataset, we utilise the full, five-band $NUV-U-B-V-I$ coverage provided by the PHANGS-HST programme, and ensure that our images are quantitatively similar to those on which our neural networks -- \texttt{VGG19-BN} \citep{SIMONYAN14} and \texttt{ResNet18} \citep{HE16} -- have been pre-trained (see Section~\ref{sec:Procedure} for further details on these networks). 

The pre-training of these networks utilises the \texttt{ImageNet}\footnote{\url{http://www.image-net.org}} dataset \citep{DENG09} consisting of 299 x 299 pixel images, which is far wider than the several pixels that the star clusters in our study typically subtend. To reduce the number of neighbouring objects within each of our training images, we extract a smaller region of 50 x 50 pixels\footnote{For the range of galaxy distances in our sample, 50 pixels represents ~87--224 pc.}, remaining centred on the original target object. This follows the procedure of \citet{WEI20}, who also tested regions of 25 x 25 and 100 x 100 pixel regions and revealed the best overall results for the 50 x 50 regions. These are then resized to a 299 x 299 pixel area to match the pixel structure of the dataset used to pre-train our neural networks. This resizing is done via interpolation using the \texttt{resize} function from the python image processing package \texttt{scikit-image}. For each target in the sample, these cutouts, which we refer to as ``postage stamps", are produced for all five of the $NUV-U-B-V-I$ broadband images and are then stored in individual header data units (HDUs) within a single file. 

As noted in Section~\ref{sec:PriorModels}, we choose to include all available data, regardless of whether there was a detection in a particular filter. This differs from the LEGUS-based models, which changed all pixel values to zero in any filter which recorded a non-detection, and removed from training objects which recorded non-detections in multiple filters. To examine the effect this may have on model accuracy, we trained two individual models (Section~\ref{sec:Procedure}) using sources from a well-sampled test galaxy, NGC 1559. From this test, we find that the model which included all source data, regardless of detection, was overall more accurate than the model which removed the non-detection data. While Class 1, 2, and 4 accuracies were on average 5\% greater (within expected variance), Class 3 accuracy was significantly improved (+34\%) for the models which included all source data and thus validate this choice.

\subsection{Training Procedure}
\label{sec:Procedure}

For a direct comparison of our newly trained models with those presented in \citet{WEI20}, we employ the same training procedure outlined in their study. 

In our experiments, we use two neural network architectures: \texttt{VGG19-BN}, which utilises more of a standard series of convolutional layers and pooling layers feeding directly into each other \citep{SIMONYAN14}, and \texttt{ResNet18}, which utilises skip connections to pass information across layers with matrix addition to reduce the overall complexity of the network, ultimately resulting in more time-efficient training compared to \texttt{VGG} models \citep{HE16}. Both of these architectures have three input channels, so in practice, two copies are concatenated (6 total channels) in order to capture the information from all five of our broadband filters (with one channel set to constant zeros), as done in \citet{WEI20}. 

Each of these neural networks has been pre-trained with the \texttt{ImageNet} dataset \citep{DENG09}. While the \texttt{ImageNet} dataset does not feature star cluster morphologies amongst its image classifications, its power resides in its scale, diversity, and hierarchical nature. With more than 14 million images, \texttt{ImageNet} allows the \texttt{VGG19-BN} and \texttt{ResNet18} models to learn lower-level features such as shapes, curves, and edges with accurate, high-quality data. Transfer learning is then implemented by replacing the last layer of the models with randomly initialised weights, which, upon training with a new dataset, will tailor the models to higher-level features specific to that particular input data -- see \citet{GEORGE17,GEORGE18,DOMINGUEZ18,ACKERMAN18,KHAN18,BARCHI19} for examples of astronomical applications. This method of transfer learning is particularly useful when the input data set is small in comparison to the pre-training data set, as is the case presented here -- our sample of $\sim$20,000 objects is nearly 1000 times smaller than that of \texttt{ImageNet}. In our training, the pre-trained weights are provided by \texttt{PyTorch} \citep{PASZKE17}. 

To begin training, a number of objects from the training set (80\% of the overall sample) are selected, first by randomly choosing an object class and then randomly choosing an object amongst that class. This number is known as the batch size, which is 16 and 32 for the \texttt{VGG19-BN} and \texttt{ResNet18} models, respectively. These objects serve as the input and are passed through the model and have their model-predicted classes compared to their human-verified classes. The accuracy of the model is then recorded as a cross-entropy loss function\footnote{A loss function is used to evaluate and diagnose model optimisation during training. The penalty for errors in the cross-entropy loss function is logarithmic, i.e., large errors are more strongly penalised.}, which is used to determine how the weights are modified in order for the model to perform more accurately. The size of these modifications is determined by the learning rate, which we set to 10$^{-4}$---a faster learning rate will make larger modifications to the weights, which may train a model faster, but may also result in a less accurate final model. These steps are then repeated for the desired number of batches, thus 10,000 batches correspond to 10,000 modifications to the initial model. Notably, it is possible to over-train the model, where the model becomes over-specified to classify the objects in the training sample and results in poorer accuracy for classifying objects outside of the sample. 

Upon training one \texttt{ResNet18} and one \texttt{VGG19-BN} model for each of the samples and viewing their performance over time, we decide upon 5,000 and 10,000 batches for all \texttt{ResNet18} and \texttt{VGG19-BN} models, respectively. With the aforementioned batch sizes, this means that each of the neural networks is exposed to 160,000 images during the training process, which is relatively large compared to our sample size. To reduce the number of identical images presented to the networks, when an object is selected in a batch, it is randomly rotated anywhere between 0 and 360 degrees, and also has a 50\% chance of being reflected about its axes after it is resized. Thus, it is rare for the model to train on the exact same array of pixel values (i.e., image) multiple times.

For an accurate representation of performance, 10 models, each with its own unique randomly-initialised weights in its final layer, are independently trained for each sample and architecture used. For the distance-independent sample, we train 10 models using the \texttt{ResNet18} architecture and 10 models using the \texttt{VGG19-BN} architecture. However, for each of the three distance-dependent samples, we choose to train 10 models using the \texttt{VGG19-BN} architecture only. This choice is primarily motivated by the relatively similar performance of the two architectures found for both our own distance-independent models and the models presented in \citet{WEI20}. With the training of 10 models for each sample and architecture, we are able to present their mean accuracies along with their standard deviations, which are presented in the following sections.

Notably, the training and evaluation of each model in this study have been completed using Amazon Web Services (AWS), utilising their EC2 p3.2xlarge instance. For additional details on deep transfer learning and its statistical foundations, we refer the reader to \citet{WEI20}.

\section{Results}
\label{sec:Results}

Here we present the results of the training experiments described in Section~\ref{sec:Experiments}. 

\subsection{Distance-Independent Models}
\label{sec:DImodels}

\begin{figure*}
\includegraphics[width =1.0\textwidth]{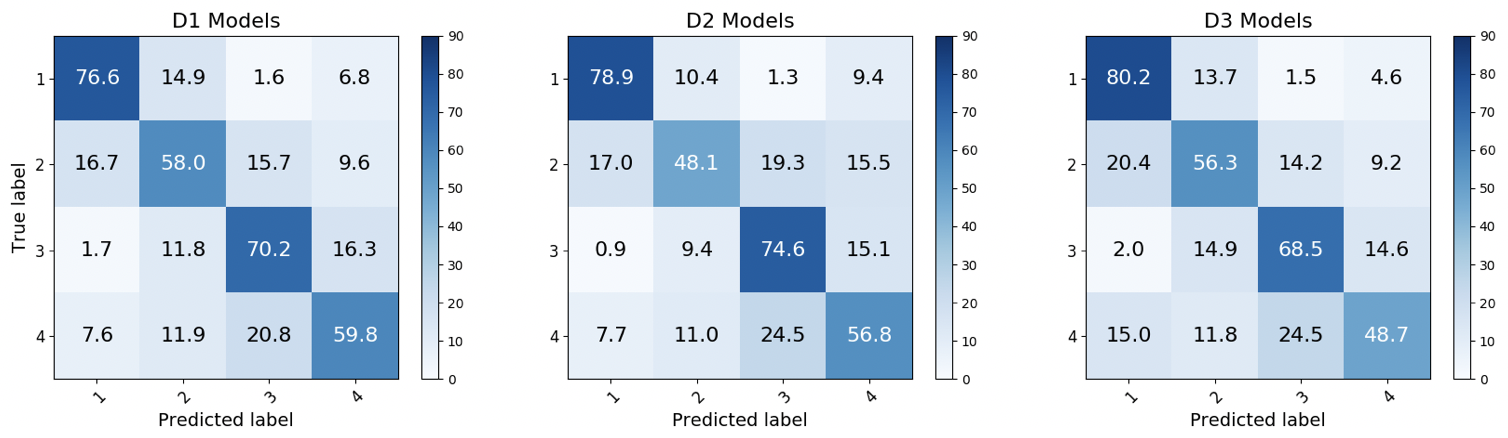}
\caption[Classification accuracy of distance-dependent models]{Confusion matrices showing the results of training separate machine learning models based on the galactic distance of PHANGS-HST objects. From left to right, the three sets of models are based on clusters between 1) 9--12 Mpc, 2) 14--18 Mpc, and 3) 18--24 Mpc. The averaged accuracies, as determined by classifying their respective validation set of PHANGS-HST objects, of 10 VGG-based models within each of the three distance bins are displayed.}
\label{fig:DisDep_results}
\end{figure*}

\begin{figure*}
\includegraphics[width =1.0\textwidth]{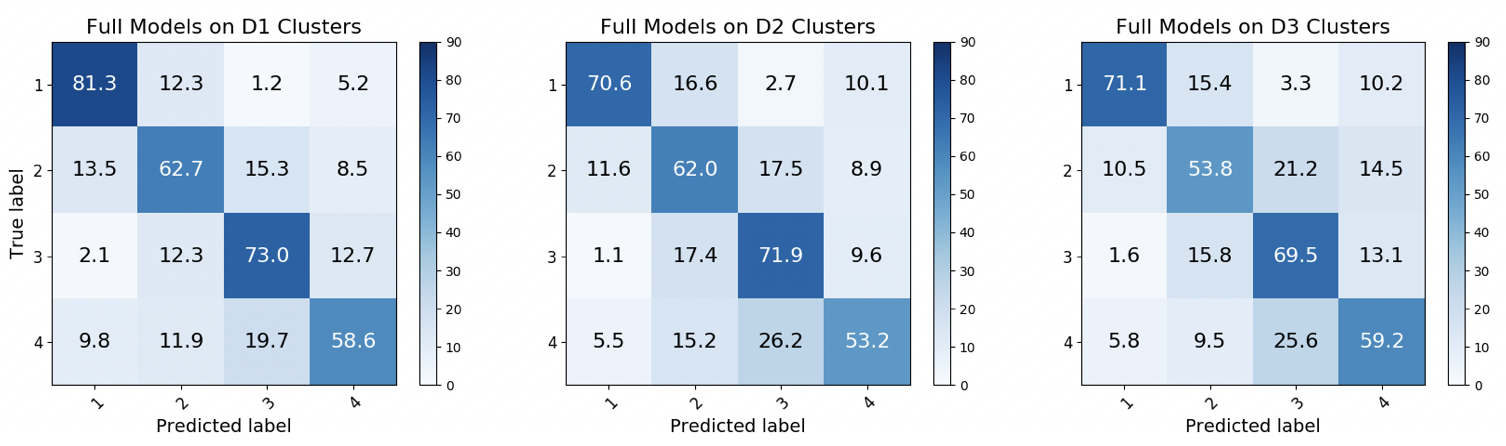}
\caption[Classification accuracy of distance-independent models in classifying the distance-dependent samples]{Confusion matrices showing the classification accuracies of our distance-independent models when classifying objects within the three distance bins shown in Figure~\ref{fig:DisDep_results} (9--12 Mpc, 14--18 Mpc, 18--24 Mpc). The similarity between these confusion matrices and those shown in Figure~\ref{fig:DisDep_results} indicates that training within distance bins does not significantly affect the accuracy of the classification models.}
\label{fig:DisInd_on_Bins}
\end{figure*}

Considering the relatively poor accuracy of the LEGUS-based models in classifying PHANGS-HST clusters (Section~\ref{sec:PriorModels}), we first train a set of new models using only data from the PHANGS-HST sample, independent of galaxy distance. The results of this training are displayed in the bottom-right confusion matrix in Figure~\ref{fig:Wei_results}, which measures the accuracy of our models in the classification of our validation set of clusters (Section~\ref{sec:Data}).  Figure~\ref{fig:Wei_results} shows the percentage of objects from each of the human-verified classes (y-axis) that receive a specific classification as predicted by the \texttt{VGG19-BN} models (x-axis). Thus, the diagonal in these plots represents clusters that received the same predicted class as their human-verified class. We remind the reader that these accuracies are based on the classifications of our validation set, the objects of which are not included in the training of the models. Equivalent confusion matrices for the \texttt{ResNet18} models are not included because of their similarity to the \texttt{VGG19-BN} results, but are discussed below.

Overall, we find marked improvement for our models over the LEGUS-based models in classifying PHANGS-HST objects, particularly for Classes 2 and 3. The accuracies averaged over our 10 \texttt{VGG19-BN} models (those presented in Figure~\ref{fig:Wei_results}), are 74 $\pm$ 10\%, 59 $\pm$ 12\%, 71 $\pm$ 9\%, and 57 $\pm$ 5\%  for Class 1, 2, 3, and 4 objects, respectively, and the accuracies averaged over our 10 \texttt{ResNet18} models (not shown in Figure~\ref{fig:Wei_results}) are 77 $\pm$ 7\%, 58 $\pm$ 10\%, 70 $\pm$ 5\%, and 62 $\pm$ 6\%. 

These accuracies are consistent with those presented in prior works as well. \citet{WEI20} reported 71--76\%, 54--64\%, 57--60\%, and 69\% accuracy for Class 1, 2, 3, and 4 objects using \texttt{VGG19-BN} models to classify LEGUS objects, and 76--78\%, 54--58\%, 58--60\%, and 66--71\% accuracy using \texttt{ResNet18} models. \citet{PEREZ21} report similar or slightly lower accuracies for their LEGUS-based sample of cluster objects, with recovery rates of 78\%, 55\%, 45\%, for Class 1, 2, and 3 objects, while their non-cluster (Class 4) accuracy is higher at 82\%. 

Additionally, we use our PHANGS-based models to classify LEGUS objects (Figure~\ref{fig:Wei_results}; bottom-left), upon which we find model accuracies of 54\%, 59\%, 76\%, and 38\% for Class 1, 2, 3, and 4 objects respectively. Compared to using the LEGUS-based models to classify PHANGS-HST objects, this represents considerable improvement for Class 2 (27\%) and 3 (33\%) accuracy. Additionally, 40\% of LEGUS Class 1 objects are instead labelled Class 2, meaning 94\% (54\% + 40\%) of LEGUS Class 1 objects are still considered a standard cluster (either Class 1 or 2). Similarly, 76\% (38\% + 38\%) of all LEGUS Class 4 would remain outside of standard samples. With these considerations, our PHANGS-based models are more reliable in classifying LEGUS objects than vice versa.

Most importantly, our newly trained models are classifying PHANGS-HST objects with much higher accuracy and greater precision than the LEGUS-based models. The new models identify Class 2 and 3 PHANGS-HST objects with greater than twice the accuracy of the LEGUS-based models, and Class 1 PHANGS-HST objects are also identified with $\sim$12\% greater accuracy (Figure~\ref{fig:Wei_results}). These accuracies represent a statistical measure known as recall, which is defined as the fraction of relevant objects which are retrieved (e.g. the fraction of correctly identified Class 1 objects amongst all human-labelled Class 1 objects). Other statistical measures also support our new models, including precision, which is the fraction of retrieved objects which are relevant (e.g. the fraction of model-classified Class 1 objects which are also human-classified as Class 1), and F-score, which is the harmonic mean of recall and precision. For these statistics, our models classify PHANGS-HST objects with 8\% greater precision and $\sim$0.14 greater F-scores on average across the four classes compared to the LEGUS-based models.



These results indicate that our new models are capable of classifying cluster morphology with accuracies that are comparable to previous machine learning studies \citep{WEI20,PEREZ21} as well as human-to-human variation ($\sim$70\% agreement; \citealt{WHITMORE21}). The new models constitute a great improvement to previous machine learning studies in terms of classification accuracy for Class 3 clusters. In addition, they outperform the previous LEGUS-based models of \citet{WEI20} in the classification of PHANGS-HST objects. Part of this improvement is likely due to sample consistency, as each set of models are trained on objects from a mostly unique set of galaxies. Our PHANGS-based models classify PHANGS clusters more accurately than the LEGUS-based models ($\sim$68\% vs. $\sim$40\% on average for Class 1, 2, and 3 objects -- i.e. clusters; see Figure~\ref{fig:Wei_results}), while the LEGUS-based models, on average, classify LEGUS clusters a bit more accurately than our models ($\sim$65\% vs. $\sim$63\%). Else, these improvements can be attributed to the new training, as we use the same neural network architectures. As noted in Section~\ref{sec:PriorModels}, the main updates in our training include a more specific Class 3 definition, the inclusion of non-detection data, as well as utilizing a larger sample of objects. Overall, we find that our models will be able to provide the most reliable machine learning classifications for cluster morphology for the PHANGS-HST sample, and can serve as a unique set of classification models moving forward.


\subsection{Distance-Dependent Models}
\label{sec:DDmodels}

\begin{figure*}
\includegraphics[width =1.0\textwidth]{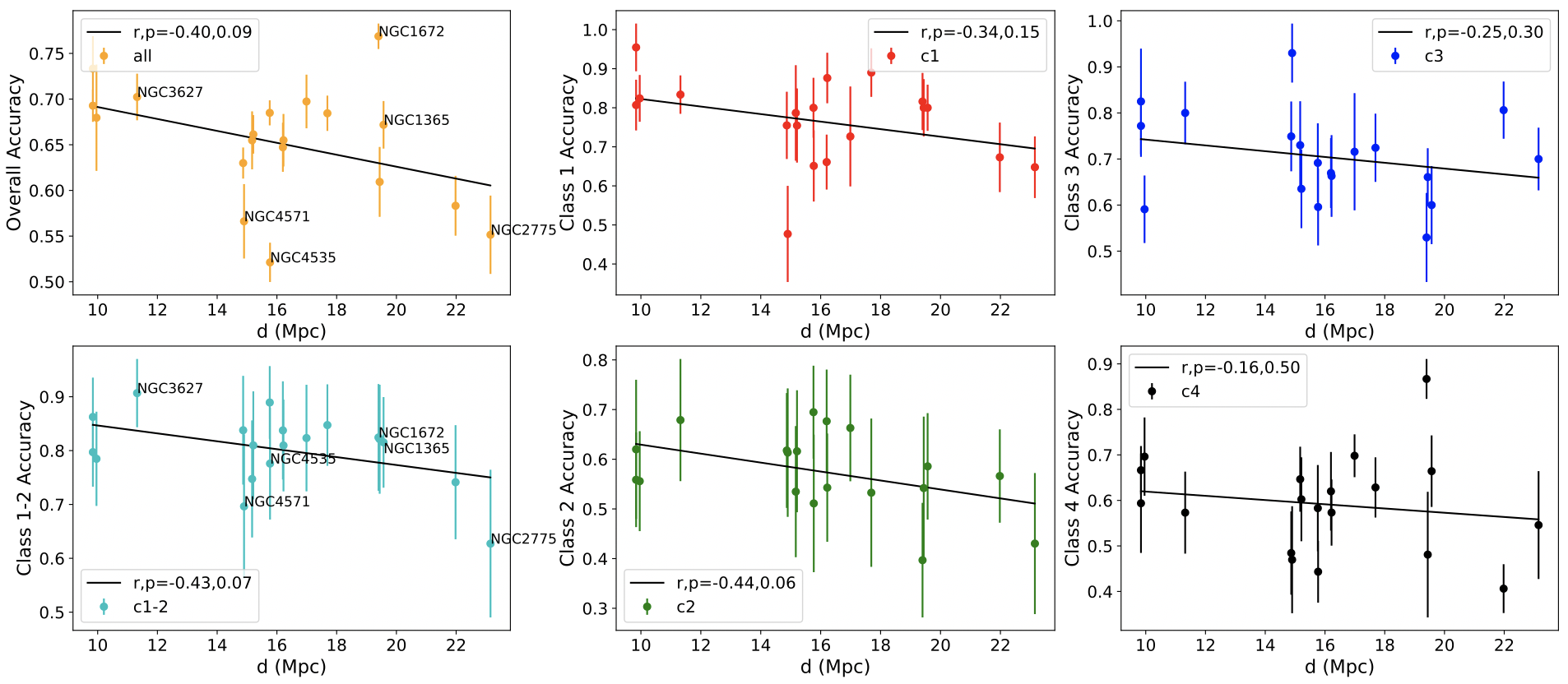}
\caption[Classification accuracy vs. galactic distance]{Classification accuracy vs. galactic distance \citep{ANAND20}. Each of the points in these plots represents the prediction accuracy based on our distance-independent models, averaged together, for objects within a particular galaxy, and the standard deviations in accuracy across the models are indicated by vertical error bars. The top-left plot shows the percent of all objects (i.e. Class 1, 2, 3, and 4) that receive identical model- and human-determined classes. The bottom-left plot shows the percent of clusters that are classified as either Class 1 or 2 by both model and human. The four plots on the right show the percentage of clusters that receive the same model- and human-determined classification for each of the four classes individually. Each plot includes a linear regression model with the Pearson correlation coefficients ($r$) and $p$-values included in the legend for reference. While model accuracy appears to generally decline with galactic distance, the correlation is not statistically significant.}
\label{fig:Acc_v_Distance}
\end{figure*}

With the improved accuracy of our new models and the fact that the LEGUS-based models are trained on a cluster sample which spans a different galactic distance range ($\sim 4-10$ Mpc) than the PHANGS-HST cluster sample used in this study ($\sim 9-24$~Mpc), it is fair to question whether the classification accuracy of these machine learning models depends on the galaxy distance. One way to examine this is to split our sample of galaxies into three separate distance bins and independently train three sets of models to allow us to compare the performance of each.

Upon dividing the sample into our three distance bins (Section~\ref{sec:Experiments}), each of our three resultant samples contains between 5,000 and 11,000 objects, which is comparable in size to the samples in previous star cluster classification studies (5,000--15,000 objects; \citealt{WEI20,PEREZ21}). For this experiment, we choose to train models using only the \texttt{VGG19-BN} architecture due to the very similar performances observed in the training of our distance-independent models (Section~\ref{sec:DImodels}).

The results of this training are displayed in the confusion matrices in Figure~\ref{fig:DisDep_results}, where the left, middle, and right plots show the accuracies of the 9--12~Mpc (D1 model), 14--18~Mpc (D2 model), and 18--24~Mpc (D3 model) bins, respectively.

Overall, we find that the accuracies of the three sets of models are consistent both with each other and with the distance-independent models. The accuracies averaged over the 10 \texttt{VGG19-BN} models for the three distance bins range from 77--80\%, 48-58\%, 68--75\%, and 49--60\% for Class 1, 2, 3, and 4 objects, respectively. If we take the averaged accuracy across all four classes, we find that accuracy slightly decreases with increased distance, though within our accuracy uncertainty. The averaged accuracy for the D1 models is 66.2\% compared to 64.6\% and 63.4\% for the D2 and D3 models, respectively. Most of the individual agreement fractions also fall within the standard deviations of the distance-independent \texttt{VGG19-BN} models. The only exception is Class 4 accuracy for the 18--24~Mpc models, which show a higher percentage of objects being reclassified as Class 1 objects instead, potentially raising a concern in the cleanness of the resulting star cluster sample. 

Additionally, we find that the distance-independent models perform similarly when classifying the validation sets from each of the three distance bins. Figure~\ref{fig:DisInd_on_Bins} displays the results of this validation testing, where we find that the distance-independent models classify clusters within the three distance bins at accuracies of 71--81\%, 54--63\%, 69-73\%, and 53--59\% for Class 1, 2, 3, and 4 objects, respectively. While the averaged accuracy across all four classes slightly drops from 68.9\% for the D1 sample to 64.4\% and 63.4\% for the D2 and D3 samples, respectively, these are again within our accuracy uncertainties. We find that these results are consistent with both the distance-dependent models (Figure~\ref{fig:DisDep_results}) as well as the accuracy of the distance-independent models when tested on the overall sample (right matrix of Figure~\ref{fig:Wei_results}).

Thus, we find that training separate models based on the galactic distance of the objects does not significantly affect their performance relative to the distance-independent models. Because of this, we determine that it is best to use the distance-independent models in the production-scale classification of PHANGS-HST objects, which is not only simpler, but will also help to avoid potential artificial correlations or discontinuities between the distance bins.

\section{Additional Trends}
\label{sec:Trends}

The accuracies of our new models, presented in Section~\ref{sec:DImodels}, are determined by using them to classify our validation set of cluster candidates, which consists of a randomly selected 20\% of the overall sample (Section~\ref{sec:Experiments}). For each of these classified objects, we retain all of its information from the PHANGS-HST cluster catalogues including its photometric data and host galaxy, which allows us to investigate more potential correlations between the performance of the models and the properties of the objects themselves as well as the properties of their host galaxies. 

\subsection{Galactic Trends}
\label{sec:GalacticTrends}


\begin{figure*}
\includegraphics[width =1.0\textwidth]{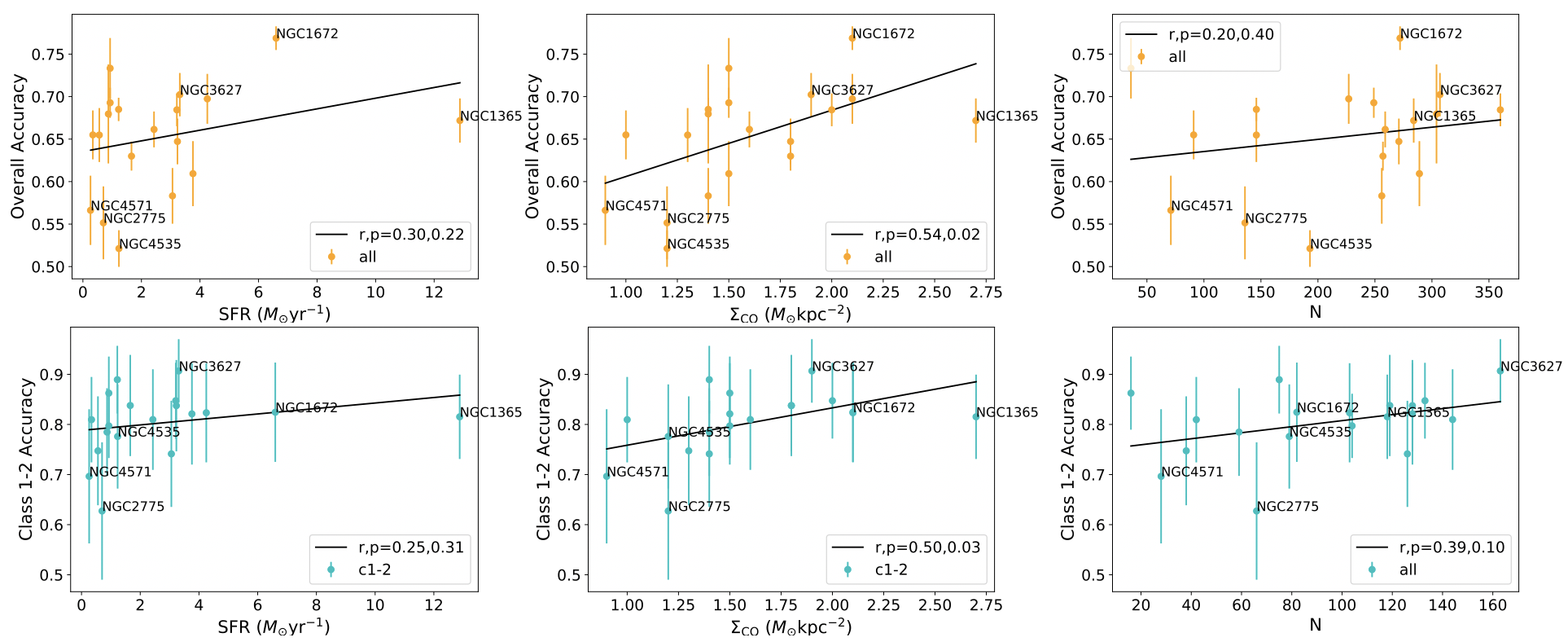}
\caption[Classification accuracy vs. SFR, $\Sigma_{\rm CO}$, and number of candidates for the field]{Classification accuracy vs. galaxy star formation rate (SFR; \citealt{SALIM16,SALIM18,LEROY21B}; left), average molecular gas surface density ($\Sigma_{\rm CO}$; \citealt{SUN18}; middle), and total number of cluster candidates in the validation set ($N$; right), based on the data provided in Table 1 of \citet{LEE21}. Vertical error bars indicate the standard deviations in accuracy across our models. Each of the points in these plots represents the prediction accuracy for objects within a particular galaxy. Each plot includes a linear regression model with the Pearson correlation coefficients ($r$) and $p$-values included in the legend for reference. Labelled galaxies are discussed in Section~\ref{sec:GalacticTrends}}.
\label{fig:Acc_v_GalParams2}
\end{figure*}

We first examine whether model performance is affected by the position or type of galaxy hosting the star clusters. The galactic properties we examine in this section have been collected from  \citet{LEROY21B} and neatly assembled in Table 1 of \citet{LEE21}. Using the distance-independent models to classify their validation set, we can then identify which of the 18 galaxies (19 fields) each object from the validation set belongs to. For each field, the validation set contains $\gtrsim$ 20 objects in each of the four classes, except for NGC~628E which only has 11, 5, 4, and 16 Class 1, 2, 3, and 4 objects accounted for, respectively.

In Section~\ref{sec:DDmodels}, we examined the performance of models that were trained in three different distance bins, but we can also analyse our model performance on a more refined scale of individual galactic distances \citep{ANAND20}, where we find a slightly negative correlation between model accuracy and galaxy distance. Figure~\ref{fig:Acc_v_Distance} displays the classification accuracy versus galaxy distance, where each point represents a single field in the sample. Also included in each of these plots is a linear regression model, including its Pearson coefficient ($r$) and $p$-value to examine statistical significance, as well as a few galaxy labels for discussion. 

Whether we look at the overall agreement regardless of class (top left plot of Figure~\ref{fig:Acc_v_Distance}), the agreement that an object is either Class 1 or 2 (bottom left plot), or the agreement for the individual classes (the remaining four plots), the lines of best fit appear to show that model accuracy declines as galaxy distance increases; however, this correlation is not found to be statistically significant. While the overall accuracy and Class 1-2 accuracy (left plots) approach significance (p-values of 0.09 and 0.07, respectively), each of the p-values are ultimately above the commonly-used threshold used to reject the null hypothesis (p < 0.05). This result is complementary to our finding that there were no obvious correlations when we trained our models in the more granular distance bins (Section~\ref{sec:DDmodels}).

\begin{figure*}
\includegraphics[width =1.0\textwidth]{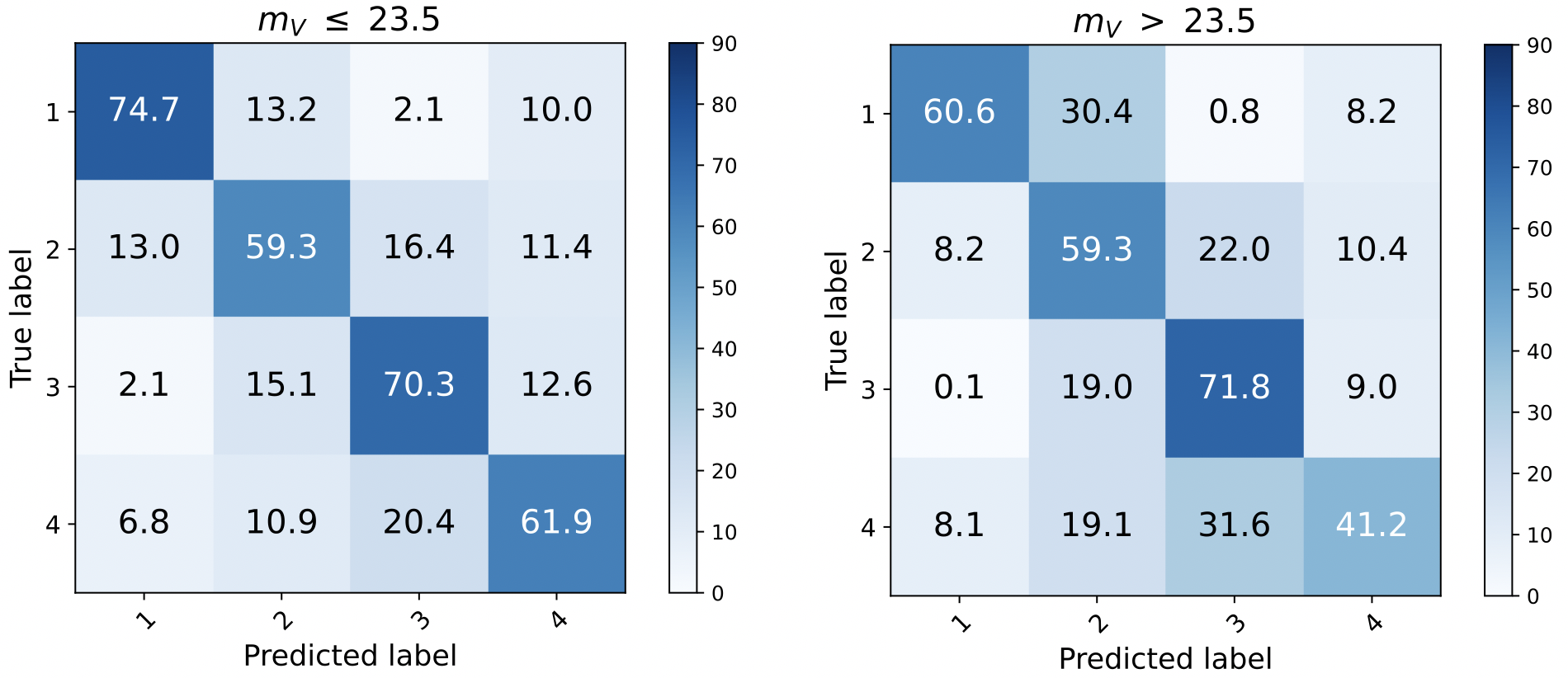}
\caption[Classification accuracies of objects based on $m_V$]{Classification accuracies of objects based on $m_V$ (VEGA). The left matrix displays model accuracies for objects with $m_V$ brighter than 23.5 mag, while the right matrix displays model accuracies for objects fainter than 23.5 mag.}
\label{fig:MagLimit}
\end{figure*}

In addition to the distance of each galaxy, the data provided in Table~1 of \citet{LEE21} allows us to investigate our model accuracy versus other galactic parameters, namely star formation rate (SFR; \citealt{SALIM16,SALIM18,LEROY21B}) and molecular gas surface density ($\Sigma_{\rm CO}$; \citep{SUN18}). Figure~\ref{fig:Acc_v_GalParams2} displays the relationships between model accuracy and each of these properties for each of the galaxies in our sample. In each of these figures, we include plots for the overall accuracies regardless of cluster class (top row), as well as for the agreement that an object is Class 1 or 2 (bottom row). While data for each galaxy's stellar mass and inclination angle are also provided in \citet{LEE21}, we did not observe any notable trends between them and model accuracy and thus do not include them here.

We identify a relatively weak, but statistically significant positive correlation between model accuracy and $\Sigma_{\rm CO}$ ($r,p = 0.54, 0.02$). Furthermore, while there is no statistically significant trend between accuracy and SFR, three of the galaxies with the highest SFR (NGC 1365, NGC 1672, NGC 3627; labelled in Figures~\ref{fig:Acc_v_Distance} \& \ref{fig:Acc_v_GalParams2}), which also have the highest $\Sigma_{\rm CO}$, have three of the highest overall accuracies, regardless of distance (Figure~\ref{fig:Acc_v_Distance}). Similarly, three of the galaxies with the lowest SFRs and $\Sigma_{\rm CO}$ (NGC 2775, NGC 4571, NGC 4535; also labelled in Figures~\ref{fig:Acc_v_Distance} \& \ref{fig:Acc_v_GalParams2}) have the three lowest overall classification accuracies. 

However, these observations may be an artefact of our sample demographics, as these galaxies with higher SFRs and $\Sigma_{\rm CO}$ naturally contain more cluster candidates (right column of Figure~\ref{fig:Acc_v_GalParams2}), hence the models are exposed to more objects within those galaxies. Furthermore, the three galaxies mentioned above with higher SFRs (NGC~1365, NGC~1672, NGC~3627) have higher percentages of Class 1 objects (Table~\ref{tab:sample}), which the models are more accurate in classifying (Figure~\ref{fig:Wei_results}), whereas the other three galaxies (NGC~2775, NGC~4571, NGC~4535) have a higher percentage of Class 2 and 3 objects, for which the models generally perform worse (Figure~\ref{fig:Wei_results}).

Overall, while it appears that model accuracy slightly decreases with individual galaxy distance, the correlation is not statistically significant, and is thus consistent with our findings for the distance-dependent models (Section~\ref{sec:DDmodels}). Additionally, while we find that our model accuracy improves for galaxies with higher $\Sigma_{\rm CO}$, it may in fact be a result of bias in the training sample of clusters.

\begin{figure*}
\includegraphics[width =0.75\textwidth]{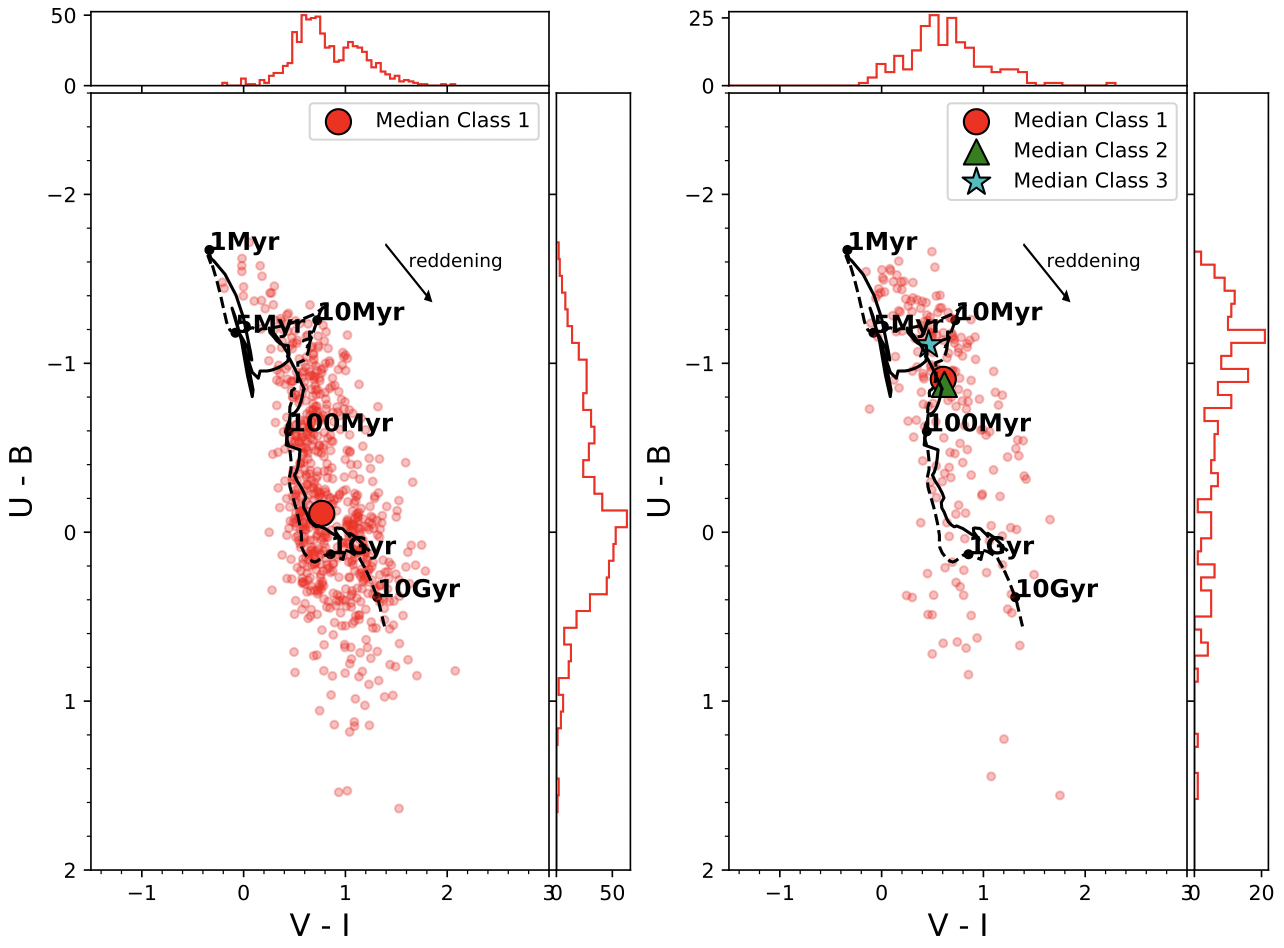}
\caption[($U-B$) vs. ($V-I$) diagrams comparing clusters for which human and machine learning classifications agree and disagree]{($U-B$) vs. ($V-I$) plots comparing clusters for which human and machine learning classifications agree (left) and disagree (right). All of the clusters in the validation set that have Class 1 morphologies as determined by a human (BCW) are included as small red circles. The left plot contains the clusters for which the mode class from the 10 VGG19-BN models is also 1, while the right contains the clusters that have a mode Class of 2, 3, or 4. \citet{BRUZUAL03} model tracks (dashed line for $Z=0.02$; solid line for $Z=0.004$) used to fit these clusters are included with time stamps for reference. The median colour for each sample is included as a larger, black-outlined circle, and histograms showing the distributions of colours are shown on each axis. Classification agreement appears to be higher for clusters which share the same colour space as old, globular clusters (e.g. the median colour in the left plot, near the 1~Gyr point), and is lower for objects sharing colour space with younger clusters (e.g., the median colour in the right plot, near the 50~Myr point) which is more consistent with the median colours of BCW Class 2 and 3 objects (denoted by a green triangle and blue star, respectively).}
\label{fig:CC_class1}
\end{figure*}

\subsection{Individual Cluster Trends}
\label{sec:IndividualTrends}

In addition to the galactic trends discussed in the previous section, we are also able to investigate the accuracy of the models based on the properties of individual star clusters. 

\subsubsection{Cluster Brightness}
\label{sec:clusterbrightness}

We first examine the accuracy as a function of cluster brightness. As described in \citet{WHITMORE21}, the standard limit for star cluster classification is $m_V$ = 23.5 mag for the PHANGS-HST sample of galaxies, however classifications were made for objects as faint as $\sim$24.5~mag for testing. With this in mind, we divide our validation sample into clusters brighter than or fainter than $m_V$ = 23.5 mag. 

Figure~\ref{fig:MagLimit} shows the confusion matrices describing the model accuracy for each of these samples, where we observe a relatively poor 1:1 accuracy for fainter Class 1 and 4 objects while Class 2 and 3 objects show no clear distinction. For the \texttt{VGG19-BN} models shown in the figure, Class 1 and 4 accuracy drops from 75\% and 62\% to 61\% and 41\%, respectively, while Class 2 and 3 accuracies remain within a standard deviation of each other. The \texttt{ResNet18} models perform similarly, with the Class 1 and 4 accuracy dropping from 79\% and 65\% to 59\% and 50\%, respectively.

Although we do not observe the decline in Class 2 accuracy with cluster brightness as shown in \citet{WHITMORE21}, we do observe their other primary result from their equivalent analysis: when we consider Classes 1 and 2 together as a single class, the model accuracy is similar on both sides of the limit. For example, from the left matrix of Figure~\ref{fig:MagLimit}, 88\% of brighter ($m_V$ $\leq$ 23.5), human-verified Class 1 clusters are classified by our \texttt{VGG19-BN} models as either Class 1 or 2 (75\% + 13\% = 88\%). Similarly, 91\% of fainter ($m_V$ > 23.5), human-verified Class 1 clusters are classified by our \texttt{VGG19-BN} models as either Class 1 or 2 (61\% + 30\% = 91\%). That is, the drop in accuracy for fainter Class 1 objects is due to more of them being classified as Class 2 clusters instead, and not as a Class 3 or 4. Similarly, the drop in Class 4 accuracy is mainly attributed to these objects being reclassified as Class 3 instead, indicating that the models have a more difficult time distinguishing between Classes 1 vs. 2 and Classes 3 vs. 4 for fainter objects likely due to the relative increase in noise. This is an important distinction because Class 1 and 2 clusters together typically represent the standard sample in star cluster studies (Section~\ref{sec:Data}), and so such samples would be unaffected by whether an object is identified as Class 1 or 2. Together, the Class 1 and 2 accuracy is $\sim$85\% for both sets of models and both brightness bins, which is consistent with \citet{WHITMORE21}.

\subsubsection{Colour-Colour Diagrams}
\label{sec:CC}

Colour-colour diagrams offer another useful tool for analysing star clusters, particularly because they allow us to view them in relation to the single stellar population (SSP) model used for the fitting of their spectral energy distribution (SED), from which their age, $E(B-V)$, and mass can be derived. We utilise the multi-band photometry of PHANGS-HST to examine the positions of clusters in ($U-B$) vs. ($V-I$) space, as shown in Figure~\ref{fig:CC_class1}. In both of these figures, each red circle represents a cluster classified by BCW as Class 1. The left plot contains each of these clusters which also received a Class 1 label from the \texttt{ResNet18} models (as determined by the mode of the 10 models), while the right plot contains the clusters that received a different classification (i.e. Class 2, 3, or 4). Also included are the colour distributions on each axis for each sample.



From these plots, we find that for Class 1 objects, the machine learning models are more likely to agree with the human-verified class for those which appear older than for those which appear younger. This is highlighted by the position of the median colours of each sample, denoted by the larger, black-outlined circles. Clusters that have been labelled Class 1 by both BCW and the models (left plot) have median colours of $U-B \approx -0.2$ and $V-I \approx 0.8$, near the 1 Gyr point of the $Z=0.02$ SSP model. This point also overlaps with the median of the overall distribution of BCW Class~1 objects. Class 1 clusters which have been incorrectly identified by the models (right plot), however, are much more concentrated toward the younger end of the SSP models, with median colours of $U-B \approx -0.8$ and $V-I \approx 0.6$, near the 50 Myr point. 

Previous studies have also identified correlations between the colours of star clusters and their morphological class (e.g., \citealt{ADAMO17,GRASHA19,TURNER21,WHITMORE21}), and more specifically that older clusters tend to be both redder and more symmetric (i.e., Class~1-like). In fact, nearly all of the objects in our sample that occupy the section of $U-B$ versus $V-I$ space designated by \citet{WHITMORE21} for older clusters ($0.95 < V-I < 1.5; -0.4 < U-B < 1.0$) are found to have accurate classifications. Only 13 of the 206 BCW Class 1 clusters in this region (6.3\%) received a different classification from our models. Furthermore, 8 of those 13 were labelled Class 2 instead, and would thus remain in standard star cluster samples. 


While our models perform well in general for Class 1 clusters, and in particular for older-appearing clusters, the distribution in the right plot of Figure~\ref{fig:CC_class1} shows that the relatively few problems they do have tend to be with younger-appearing clusters. With the median colour near the 50~Myr point in the SSP model, BCW Class 1 clusters that have different model classifications are generally much younger looking than those that agree (median near 1~Gyr). Searching the \citet{WHITMORE21} box designated for the youngest objects ($-0.65 < V-I < 0.95; -1.8 < U-B < -1.1$), we find that 58 of 117 ($\sim$50\%) of the BCW Class 1 clusters  received a different classification from our models. These younger-appearing Class 1 objects share the same colour space as Class 2 and 3 objects (highlighted in the right plot of Figure~\ref{fig:CC_class1}), which have median colours of $U-B \approx -0.8$, $V-I \approx 0.6$ and $U-B \approx -1.1$, $V-I \approx 0.5$, respectively. What is reassuring again is that the majority of the reclassified Class 1 objects (34 of the 58) are determined to be Class 2 instead. Thus, even in this relatively problematic region, $\sim$80\% of Class 1 objects would still be retained in standard star cluster catalogues. 

These results indicate that our neural network models appear to take colour into account in the classification of star clusters, where Class 1 objects are more accurately classified when they appear redder in colour. Importantly, the distributions in Figure~\ref{fig:CC_class1} indicate that this result may stem from training bias -- aligned with expectation, there is a higher frequency of bluer Class 2 and Class 3 objects versus a higher frequency of redder Class 1 objects. 

Lastly, equivalent plots for Class 2, 3, and 4 objects are included in Figure~\ref{fig:CC_234} of Appendix A, however the color distributions between correctly-classified and misclassified objects are not clearly distinguishable as is the case for Class 1 objects. This may be related to the fact that the colour distributions of Class 2, 3, and 4 objects appear much smoother than the distribution of Class 1 objects, which is more clearly double-peaked at the younger and older ends of the SSP model. Thus, while colour may help inform the classification of star clusters, particularly for Class 1 objects, it does not appear to be a primary classification criterion for all objects.

\subsubsection{SED Ages}

\begin{figure}
\centering
\includegraphics[width =0.48\textwidth]{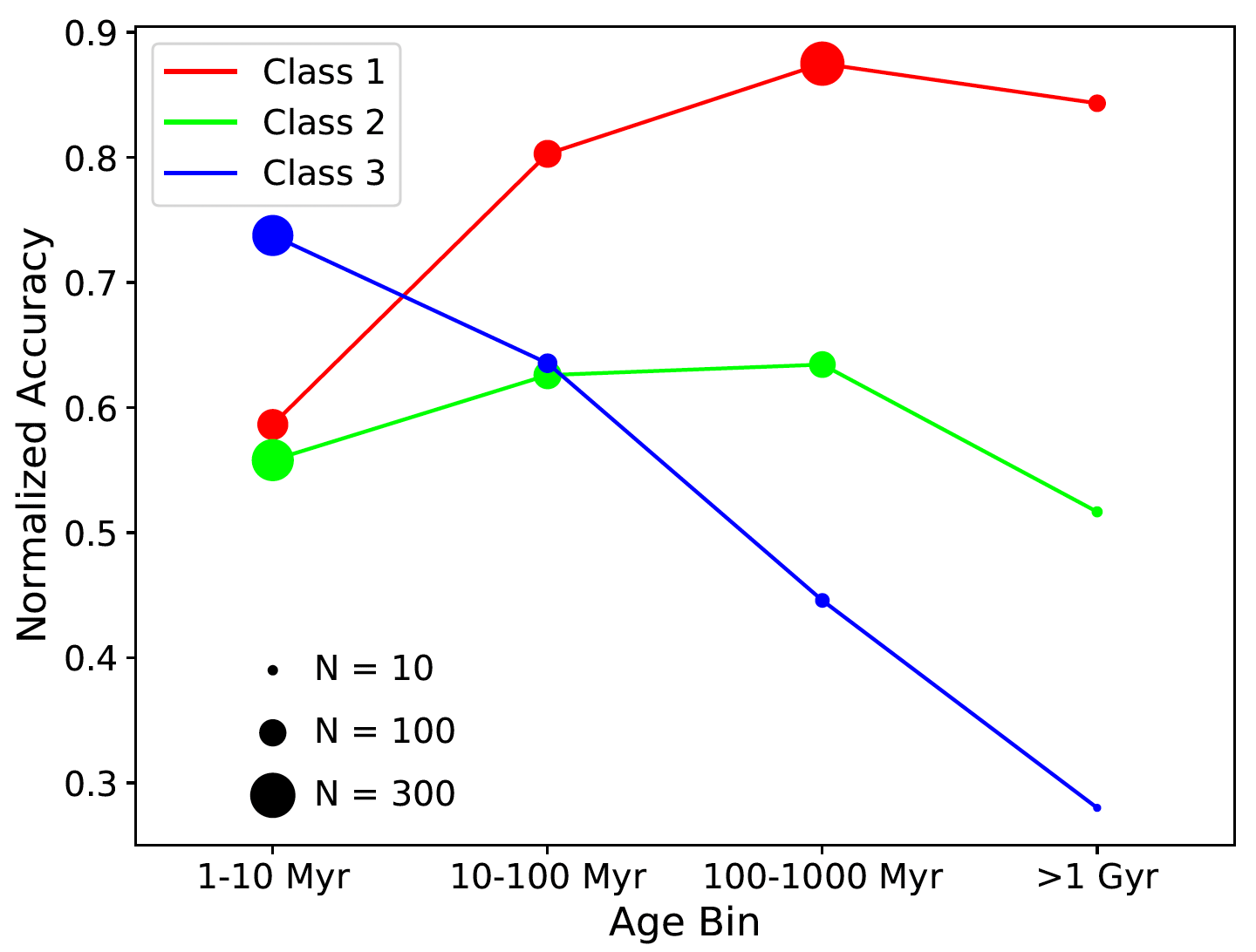}
\caption[Classification accuracy vs. SED age of clusters]{Average classification accuracies of PHANGS-HST clusters (Class 1, 2, and 3 objects), divided into four SED age bins (1-10 Myr, 10-100 Myr, 100-1000 Myr, >1 Gyr). Classification accuracy for Class 1 objects increases as the clusters become older, while accuracy decreases for Class 3 objects as they become older. The displayed normalized accuracies are based on the average of the 10 \texttt{VGG19-BN}-based distance-independent models tested on our validation set of clusters, and each point is weighted in size by the number of objects in its respective bin.}
\label{fig:age-class}
\end{figure}

The observed colour-dependence of our model accuracy is, in turn, reflected in how the model accuracy varies with the SED ages of our objects. These ages are derived by SED-fitting of the $NUV-U-B-V-I$ photometry of each object with Code Investigating GALaxy Emission\footnote{\url{http://cigale.lam.fr}} (CIGALE; \citealt{BURGARELLA05,NOLL09,BOQUIEN19}). The fitting uses the single-aged population synthesis models of \citet{BRUZUAL03} and assumes solar metallicity, a \citet{CHABRIER03} initial mass function with limits of 0.1-100 M$_{\odot}$, and a \citet{CARDELLI89} extinction curve with $R_{\rm V}$ = 3.1 (see \citealt{TURNER21} for more details on the SED-fitting of PHANGS-HST objects). Figure~\ref{fig:age-class} displays the average model accuracies for each of the three classes of clusters (Class 1, 2, and 3), divided into four distinct age bins. 

As discussed in Section~\ref{sec:CC}, our models more accurately identify Class 1 objects which appear older, however not just in colour-space, but also according to their SED age. Class 1 accuracy is lowest for the youngest objects (1-10 Myr; $\sim$58\%), and improves dramatically for the older objects: both the 100-1000 Myr and >1 Gyr age bins have Class 1 accuracies $\gtrsim$ 85\%. Additionally, we observe the opposite effect for Class 3 objects, namely that as SED age increases, Class 3 accuracy decreases: accuracy of these objects is best for the youngest clusters (71\%) and dramatically \textit{decreases} as they become older (28\% accuracy for those with SED ages >1 Gyr). Class 2 objects do not reveal such obvious correlations. These results are complementary to those discussed in Section~\ref{sec:CC}, and again may be a reflection of training bias. As shown by the weighted points in Figure~\ref{fig:age-class}, there is a higher frequency of young (bluer) Class 2 and Class 3 objects versus a higher frequency of older (redder) Class 1 objects.

It should be noted that while the models may take the colour (and thus indirectly the age) of these clusters into account, the distributions in Figure~\ref{fig:CC_class1} and the weighted points in Figure~\ref{fig:age-class} indicate that these results may also be a result of training bias -- there is a higher frequency of young Class 2 and Class 3 objects versus a higher frequency of older Class 1 objects.

Overall, we observe reduced agreement between the BCW and our model classes for objects fainter than the standard magnitude limit ($m_V$ > 23.5, particularly for Class 1 and 4) as well as those that appear younger in the $U-B$ vs. $V-I$ space (Class 1), while also noting that the models correctly label BCW Class 1 objects which occupy the same colour space as old, globular clusters with a very high degree of accuracy. This matches what we observe with the SED ages of objects as well, where we also find that Class 3 objects are most accurately classified for the youngest objects instead, though this may be a reflection of the sample and training distributions. Importantly, the instances where we do observe reduced accuracy for Class 1 objects are mostly disagreements between a Class of 1 or 2 label, and when we consider the two classes together as many star cluster studies do, then the overall accuracy remains good ($\gtrsim$ 80\%), consistent with our previous findings. 



\section{Summary}
\label{sec:Summary}

We present the results of new neural network models trained with deep transfer learning techniques for star cluster morphological classification using catalogues from the PHANGS-HST survey for 23/38 galaxies. Our sample consists of more than 20,000 star cluster candidates, all of which have been visually inspected by one of the authors (BCW). We utilise the standard cluster morphology classification system which has been used across LEGUS and PHANGS-HST studies among others, and includes four classes: 1) symmetric, 2) asymmetric, 3) multi-peaked, 4) non-cluster (e.g. single star or background galaxy). 

Our primary experiments use the \texttt{ResNet18} and \texttt{VGG19-BN} neural network architectures, and transfer their pre-trained \texttt{ImageNet} knowledge for the task of classifying cluster morphology for the PHANGS-HST sample. In our first experiment, we use 80\% of the available PHANGS sample (independent of galactic distance) to train 10 \texttt{ResNet18} models and 10 \texttt{VGG19-BN} models, and evaluate their performance by using them to classify the remaining 20\% of clusters in the sample. Our second experiment then divides the cluster sample into three separate bins based on galactic distance (9--12 Mpc, 14--18 Mpc, 18--24 Mpc). These three samples are used to independently train three sets of 10 models using the \texttt{VGG19-BN} architecture (found to have consistent results with \texttt{ResNet18} during the initial experiment). We refer to these as our distance-independent and distance-dependent models, respectively.

The results of these experiments are summarised as follows.

\begin{enumerate}
    \item Our new models show considerable improvement over the previous LEGUS-based models of \citet{WEI20} when used to predict the classes of PHANGS-HST objects. Our new models classify Class 1, 2, 3, and 4 objects with accuracies of $\sim$74\%, $\sim$59\%, $\sim$71\%, and $\sim$57\%. This represents improvements of $\sim$12\%, $\sim$32\%, and $\sim$38\% over the LEGUS-based models in the classification of PHANGS-HST Class 1, 2, and 3 objects (i.e., star clusters). 
    
    \item The division of the training set sample into three separate distance bins does not significantly affect the accuracy of the models. The accuracies averaged over the 10 \texttt{VGG19-BN} models for the three distance bins range from 77--80\%, 48--58\%, 68--75\%, and 49--60\% for Class 1, 2, 3, and 4 objects, respectively. We have used our distance-independent models to classify each of the corresponding distance-binned validation sets as well, with accuracies ranging from 71--81\%, 54--63\%, 70--73\%, and 53--59\% for Class 1, 2, 3, and 4 objects, respectively. 
    
The results of these experiments have informed the decision to use the distance-independent models for the production-scale classification of the full set of cluster candidates in all 38 PHANGS-HST galaxies. 

    
Additionally, we investigate the dependence of model accuracy on a variety of cluster and galaxy-host properties, upon which we find: 

    \item The model accuracy appears to slightly decrease as the galactic distance increases ($\lesssim$10\% from 10--23~Mpc). We also find that model accuracy is higher for galaxies with greater star formation activity, as indicated by SFR and $\Sigma_{\rm CO}$. While we note that these trends are weak, they may also be a result of demographics: the galaxies with the highest star formation rates naturally have the most clusters on which the models have been trained, and these galaxies also have a greater percentage of Class 1 objects, which the models are more adept at classifying. 
    
    \item The model accuracy for Class 1 objects is lower for faint objects as well as those which appear blue (young) in colour-colour space. The recovery rate for Class 1 objects fainter than $m_V$ = 23.5 mag is 14\% lower than for objects brighter than $m_V$ = 23.5 mag. 
    The oldest Class 1 objects (> 100 Myr) are identified by our models with $\sim$30\% greater accuracy than those which are youngest ($\leq$ 10 Myr). 
    On the other hand, for Class 3 objects the model accuracy is highest for the youngest objects (71\%), and is significantly worse for the oldest objects (28\% accuracy for objects with SED age > 1~Gyr). Importantly, these results may be a reflection of the sample and training distributions, as there is a higher frequency of younger (bluer) Class 2 and 3 objects versus a higher frequency of older (redder) Class 1 objects.
    

    We find that most Class 1 clusters are mis-classified as Class 2 clusters, which provides reassurance for the use of combined Class 1 and 2 samples, which typically form the basis of star cluster studies.
\end{enumerate}

As we find that our newly trained models are effective in classifying star clusters from PHANGS-HST, the next objective will be to generate machine learning classifications for the full sample of PHANGS-HST star cluster candidates ($\sim$200,000 sources in 38 galaxies) using these models. These classifications will be included in the next public release of the PHANGS-HST cluster catalogues at \url{https://archive.stsci.edu/hlsp/phangs-cat}.


There are also opportunities to explore enhanced model performance. For example, newer neural networks such as Contrastive Language–Image Pre-training (CLIP; \citep{RADFORD21}) offer unique algorithms which have shown improved classification accuracy over \texttt{ResNet} models for objects outside of the dataset on which the models were trained (i.e. \texttt{ImageNet}; \citealt{DENG09}). Another strategy could involve the use of different model inputs, such as a more in-depth analysis of the inclusion of non-detection data presented in Section~\ref{sec:TrainingImages}, or perhaps the use of colors rather than individual broadband data. However, continuing to optimise the performance of classification models will also depend on improvement in consistency between different human classifiers, and the development of a standardised data set of human-labelled star cluster classifications, agreed upon by a full range of experts in the field, as discussed in \citet{WEI20}.



\section*{Acknowledgements}
Based on observations made with the NASA/ESA Hubble Space Telescope for Program 15654, obtained from the MAST data archive at the Space Telescope Science Institute. STScI is operated by the Association of Universities for Research in Astronomy, Inc. under NASA contract NAS 5-26555. 

M\'ed\'eric Boquien gratefully acknowledges support by the ANID BASAL project FB210003 and from the FONDECYT regular grant 1211000. MC gratefully acknowledges funding from the DFG through an Emmy Noether Research Group (grant number CH2137/1-1). JMDK gratefully acknowledges funding from the European Research Council (ERC) under the European Union's Horizon 2020 research and innovation programme via the ERC Starting Grant MUSTANG (grant agreement number 714907). COOL Research DAO is a Decentralised Autonomous Organisation supporting research in astrophysics aimed at uncovering our cosmic origins. KG is supported by the Australian Research Council through the Discovery Early Career Researcher Award (DECRA) Fellowship DE220100766 funded by the Australian Government. KG is supported by the Australian Research Council Centre of Excellence for All Sky Astrophysics in 3 Dimensions (ASTRO~3D), through project number CE170100013. SD is supported by funding from the European Research Council (ERC) under the European Union’s Horizon 2020 research and innovation programme (grant agreement no. 101018897 CosmicExplorer).

\section*{Data Availability}
The broadband images and star cluster catalogues used in this study are taken from PHANGS-HST (\citealt{LEE21}; \url{https://phangs.stsci.edu}), which includes WFC3 (F275W, F336W, F438W, F555W, and F814W) and archival ACS (F435W, F555W, and F814W) \textit{HST} imaging. The machine learning classifications presented in this paper will be included with the release of the PHANGS-HST cluster catalogues.




\bibliographystyle{mnras}
\bibliography{sjh.bib} 




\appendix
\section{}

\begin{figure*}
\includegraphics[width =1.0\textwidth]{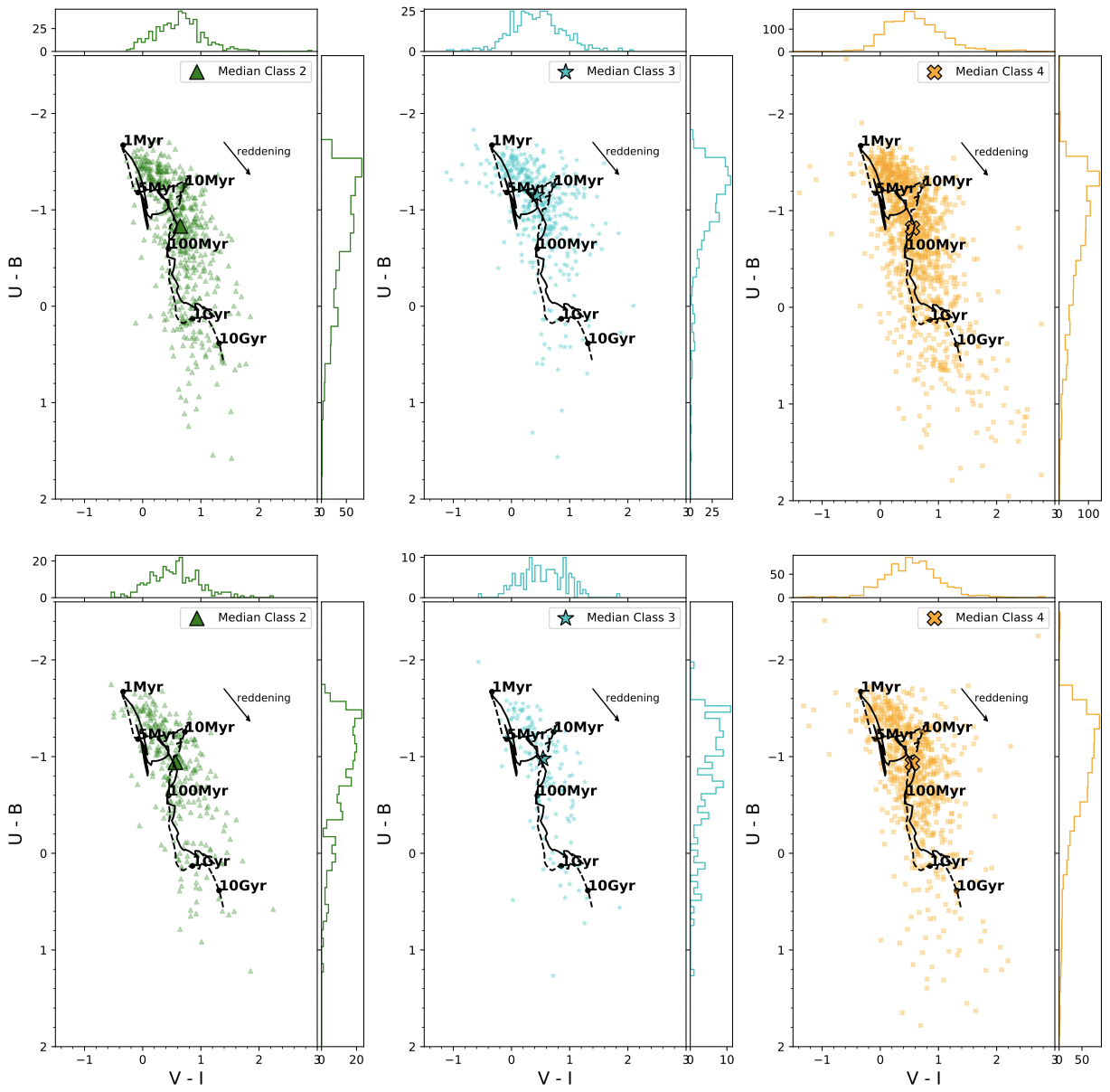}
\caption[($U-B$) vs. ($V-I$) diagrams comparing class 2, 3, and 4 sources for which human and machine learning classifications agree and disagree]{Same as Figure~\ref{fig:CC_class1} for Class 2, 3, and 4 sources. ($U-B$) vs. ($V-I$) plots comparing clusters for which human and machine learning classifications agree (top) and disagree (bottom). \citet{BRUZUAL03} model tracks (dashed line for $Z=0.02$; solid line for $Z=0.004$) used to fit these clusters are included with time stamps for reference. The median colour for each sample is included as a larger, black-outlined point, and histograms showing the distributions of colours are shown on each axis. Unlike the Class 1 clusters displayed in Figure~\ref{fig:CC_class1}, there are no significant differences in the color distributions of Class 2, 3, and 4 sources for which human and machine learning classifications agree versus disagree.}
\label{fig:CC_234}
\end{figure*}


\bsp	
\label{lastpage}
\end{document}